\documentstyle[epsfig]{mn}

\title[Foreground Contamination in CMB Maps]{Faraday Rotation as a Diagnostic of
Galactic Foreground Contamination of CMB Maps}

\author[Patrick Dineen \& Peter Coles]{Patrick Dineen \& Peter Coles\\
School of Physics \& Astronomy, University of Nottingham,
University Park, Nottingham, NG7 2RD, United Kingdom\\ }

\begin{document}

\maketitle

\begin{abstract}
We present a diagnostic test of possible Galactic contamination of
cosmic microwave background sky maps designed to provide an
independent check on the methods used to compile these maps. The
method involves a non-parametric measurement of cross-correlation
between the Faraday rotation measure (RM) of extragalactic sources
and the measured microwave signal at the same angular position. We
argue that statistical properties of the observed distribution of
rotation measures are consistent with a Galactic origin, an
argument reinforced by a direct measurement of cross-correlation
between dust, free--free and synchrotron foreground maps and RM
values with the strongest correlation being for dust and
free--free. We do not find any statistically compelling evidence
for correlations between the RM values and the COBE DMR maps at
any frequency, so there is no evidence of residual contamination
in these CMB maps. On the other hand, there is a statistically
significant correlation of RM with the preliminary WMAP individual
frequency maps which remains significant in the Tegmark et al.
Wiener-filtered map but not in the Internal Linear Combination map
produced by the WMAP team.
\end{abstract}


\section{Introduction}

The DMR (Differential Microwave Radiometer) instrument on board
the COBE satellite ushered in the era or precision cosmology by
measuring primordial temperature fluctuations of order \(\Delta
T/T=10^{-5}\) in the cosmic microwave background (CMB) radiation
\cite{smo92}.  The WMAP (Wilkinson Microwave Anisotropy Probe)
recently took CMB cosmology onto a higher level by mapping the
last scattering surface with much higher angular definition and
greater sensitivity (Bennett et al. 2003; Hinshaw et al. 2003).
But before either of these experiments could produce cosmological
information various sources of contaminating radiation had to be
eliminated.

Among the foregrounds that may affect CMB observations are
Galactic dust, synchrotron and free--free emission, extragalactic
point sources and the Sunyaev-Zel'dovich effect due to hot gases
in galaxy clusters. These fluctuations were measured by the
COBE-DMR instrument on angular scales larger than
\(\mathrm{7^{o}}\) and at 3 different frequencies (31.5, 53 and 90
GHz) corresponding to a window where the Galactic emissions are
minimum. Instruments such as COBE-DMR, which probe large angular
scales,  are most sensitive to the diffuse foreground emission of
our galaxy \cite{gs00}. Synchrotron emission associated with the
motion of electrons in the Galactic magnetic field  dominates at
\(\nu \la\) 20 GHz. Free-free emission, due to electron-ion
collisions in the interstellar medium, dominates in the range
~25-70 GHz. Galactic dust grains absorb UV and optical light from
the interstellar radiation field and emit the energy in the
far-infrared. This foreground dominates at \(\nu \ga\) 90 GHz. CMB
instruments can identify foregrounds by their spectral signatures
across a range of frequencies, making reliable foreground
subtraction feasible. WMAP was designed to make accurate
measurements of the CMB down to angular scales of $\sim 12$
arcmin; to allow efficient foreground subtraction this experiment
mapped the full sky at five widely separated frequencies from 23
GHz to 94 GHz. The quest for even better foreground prediction and
subtraction will be an important part of future CMB work as finer
details of the temperature pattern are probed with increasing
precision. This is likely to be particularly challenging for
polarization studies, in which the contamination is larger in
proportion to the CMB signal than for the temperature field
\cite{dw98,ssk02}. It is therefore extremely important to find
ways of testing that foreground modelling and subtraction has been
carried out accurately, particularly given the evidence that
exists already for unusual features on the CMB sky revealed by
WMAP (e.g. Chiang et al. 2003).

In this paper we show how it is possible to construct an
independent probe of foregrounds using measurements of Faraday
rotation along lines of sight to extragalactic objects. The layout
of the paper is as follows. In the next section we give a brief
overview of possible foreground contaminations of CMB maps, and
explain how they might be correlated with Faraday rotation
measures taken along random lines-of-sight. In Section 3, we carry
out a simple and robust test to verify that there is a link
between the strength of the sources and their location in the sky.
In Section 4 we explain our non-parametric cross-correlation
technique. In Section 5, we confirm the link between the RM values
and Galactic foregrounds by determining correlations between the
RM catalogue and maps of various contributions, including the
survey of Haslam et al. (1982) at 408 MHz, synchrotron maps
compiled by the WMAP team, as well as dust and free--free maps. We
also explore correlations between the RM-values and the
temperature field values in the DMR and WMAP data at the location
in the sky of the sources to see if there is any evidence of
residual contamination in published maps. The conclusions are
presented and discussed in Section 6.

\section{Faraday Rotation and Foregrounds}

Faraday rotation measures (RM) of extragalactic radio sources are
direct tracers of the Galactic magnetic field. When
plane-polarized radiation propagates through a plasma with a
component of the magnetic field parallel to the direction of
propagation, the plane of polarisation rotates through an angle
\(\psi\) given by
\begin{equation}
\label{psi} \psi={\cal R}\lambda^{2},
\end{equation}
where the Faraday rotation measure (RM) is denoted by the symbol
${\cal R}$ and measured in $\mathrm{rad\,m}^{-2}$ where
 \begin{equation}
\label{rm} {\cal R}=\frac{e^{3}}{2\pi m^{2}_{e}c^4}\int
n_eB_\parallel \,ds.
\end{equation}
Note that $B_\parallel$ is the component of the magnetic field
along the line-of-sight direction. The observed RM of
extragalactic sources is a linear sum of three components: the
intrinsic RM of the source (often small); the value due to the
intergalactic medium (usually negligible); and the RM from the
interstellar medium of our Galaxy \cite{bmv88}. The latter
component is usually assumed to form the main contribution to the
integral. If this is true, studies of the distribution and
strength of RM values can be used to map the Galactic magnetic
field \cite{vk75}. Even if the intrinsic contribution were not
small, it could  be ignored if the magnetic fields in different
radio sources were uncorrelated and therefore simply add noise to
any measure of the Galactic field \cite{fsss01}. In a similar
vein, the distributions of RM values have been used to measure
local distortions of the magnetic field, such as loops and
filaments, and attempts have also been made to determine the
strength of intracluster magnetic fields \cite{ktk91}. In what
follows we shall use RM values obtained from a catalogue of
extragalactic sources compiled by Broten et al. (1988) (updated in
1991) not to attempt mapping the Galactic $B$-field but to look
for statistical correlations over the whole sky.

There are various ways in which rotation measures of external
galaxies could be diagnostic of foreground contamination. The most
obvious at first sight is Galactic synchrotron. The {\em
magnitudes} of rotation measures of extragalactic sources trace
the Galactic magnetic field strength which, in turn, is correlated
with the strength of synchrotron emission resulting from the
acceleration of electrons in the Galactic magnetic field. The
emission is dependent on both the energy spectrum of the
electrons, $N(E)dE$ and the strength of the magnetic field, $B$.
For a power-law distribution of energies of the form
\begin{equation}
N(E)dE = \kappa E^{-\beta}dE,
\end{equation}
the intensity spectrum of the emitted radiation takes the form
\begin{equation} I(\nu) \propto \kappa
B_\perp^{\alpha+1}\nu^{-\alpha}. \end{equation} The normalisation constant $\kappa$ is related to the overall number density of electrons, $n_e$. The intensity
spectral index $\alpha$ is related to the index of the energy
spectrum $\beta$ via $\beta=2\alpha+1$, but $\alpha$ is expected
to vary with both position and frequency \cite{lmop87}. Radio
surveys at frequencies below 2 GHz are dominated by synchrotron
emission \cite{dw98} so these have been used to extrapolate this
contribution to the higher frequencies at which CMB instruments
operate. However, the only complete-sky survey is that of Haslam
et al. (1982) at 408 MHz. Extrapolating from such a low frequency
measurement is prone to problems with zero levels and scanning
errors. Moreover, it is known that the spectral index $\alpha$
varies by $\Delta \alpha \simeq 0.5$ \cite{dwg} which could have a
big effect on extrapolations over a large frequency range. If the
DMR or WMAP data correlate with RM measurements then this may
suggest that synchrotron emission has not been completely removed.
Our approach is intended to be complementary to the standard
extrapolation.

On the other hand, correlation with Galactic synchrotron is not
the only possible diagnostic use of rotation measure data, and may
indeed not be the most important. Note that, while the formulae
for both rotation measure (2) and synchrotron intensity (4) both
depend on $B$, the former depends on the line-of-sight component
and the latter on the perpendicular component. If the magnetic
field were disordered on a relatively small scale one might still
expect correlations to exist between the two, but it is certainly
possible to imagine field configurations that result in large
synchrotron emission but no Faraday rotation (and vice-versa). On
the other hand, although free--free emission does not rely upon
the presence of a $B$-field, it does depend on the electron
density as does the rotation measure. A correlation between RM and
free--free emission is therefore possible even if the field
configuration leads to a negligible correlation with synchrotron
emission in the observer direction. Dust may likewise be
indirectly correlated; this correlation may be further complicated
if the dust is aligned in some way with the Galactic magnetic
field.

It is not obvious {\em a priori} which of the potential
contaminants would correlate best with the measured RM values nor
what the meaning of any measured correlation would be. For
example, free--free, synchrotron and dust emission all vary
strongly with Galactic latitude. For this reason alone they are
expected to correlate with each other. A cross-correlation with RM
could therefore, on the face of it, simply be a Galactic latitude
effect. On the other hand, such a cross-correlation could instead
indicate more complex, smaller-scale spatial association between
these foregrounds. We therefore adopt an entirely empirical
approach to this question; we return to the issue of Galactic
dependence later, in Section 5.

\section{The Distribution of Rotation Measures}

Broten et al. (1988) present an all-sky catalogue of rotation
measures for 674 extragalactic sources (i.e. galaxies or quasars).
It is believed that very large RM values do not reflect the
contribution of the Galactic magnetic field, but are instead due
to the magnetic field in the source or are perhaps simply
unreliable determinations of RM \cite{rs79}. The catalogue
contained 39 sources where  $ |{\cal R}| > 300
\mathrm{\,rad\,m}^{-2}$; 25 of these are within the Galactic
``cut'' of the DMR data used in Section 5.2 and 27 are within the
Kp2 mask region of the WMAP data used later in Section 5.3. While
noting this complication, we decided nevertheless to use the
complete sample for this study. The large-valued rotation measures
would only be expected to dilute observed correlations if they
were entirely intrinsic; we comment on this later, in Section 5.

Our first point of investigation was to look at the distribution
of rotation measures across the sky. If we have a set of points on
the celestial sphere labelled with some particular characteristic
that is independent of the direction on the sphere, any subsample
selected using this characteristic should display the same
behaviour as the sample as a whole. In particular, any measure of
the spatial correlations of the subsample should have the same
form as the complete sample (scaled to take into account the
smaller sample size). Looking at the clustering characteristics of
subsamples of the Faraday rotation measure catalogue can thus
indicate whether the RM values for the sources are intrinsic or
determined by their spatial positions.

Owing to the small size of the sample available for this analysis,
a simple but robust statistic is needed. We chose the conditional
distribution of neighbour distances, $\eta(\theta)$, i.e. the
probability that there is a neighbour within an angle $\theta$ of
a given point.  This has the advantage of being closely related to
the angular two--point correlation function (Peebles 1980). The
function $\eta(\theta)$ is straightforwardly calculated by
determining the angle between the $i$th source and each remaining
source, and repeating the process for every $i$.  The resulting
angles were placed in 100 bins of equal width leading to the
distribution of \(\eta(\theta)\).

The sample was divided into two sets; those with positive RM
values (364 sources) and those with zero or negative values of RM
(310 sources).  We calculated \(\eta(\theta)\) for two subsamples:
those with positive rotation measures and those with negative
ones, giving \(\eta^{+}(\theta)\) and \(\eta^{-}(\theta)\)
respectively. The two distributions were compared through Monte
Carlo (MC) simulations in which the locations of the sources in
the sky were maintained, but the RM values were randomly
reallocated to the position of the sources. This method ensures
the underlying distribution of \(\eta(\theta)\) for real and MC
samples  was the same, but any link between the RM values and
spatial position would be severed in the MCs. We extract
\(\eta^{+}(\theta)\) and \(\eta^{-}(\theta)\) from each simulation
and constructed an average of distribution over 1000 simulations.
Since the distributions are binned it is natural to compare MC
distributions with the real data to the average distributions via
a \(\chi^2\) test, using
\begin{equation}
\label{eta}
\chi^2=\sum_{i}\frac{\left[\eta(\theta_i)-\overline{\eta}(\theta_i)\right]^2}{\overline{\eta}(\theta_i)},
\end{equation}
where \(\overline{\eta}(\theta)\) is the average distribution. The
larger \(\chi^2\) is the less likely the distribution is drawn
from the population represented by \(\overline{\eta}(\theta)\).
The \(\chi^2\)statistic obtained from the positive distributions
for the MCs and Broten et al. data were ranked (lowest value to
highest). The same was done for the statistics obtained from the
negative distributions. Thus, the higher the rank of the real data
the more likely the rotation measures and positions are
correlated.

\begin{figure}
\epsfig{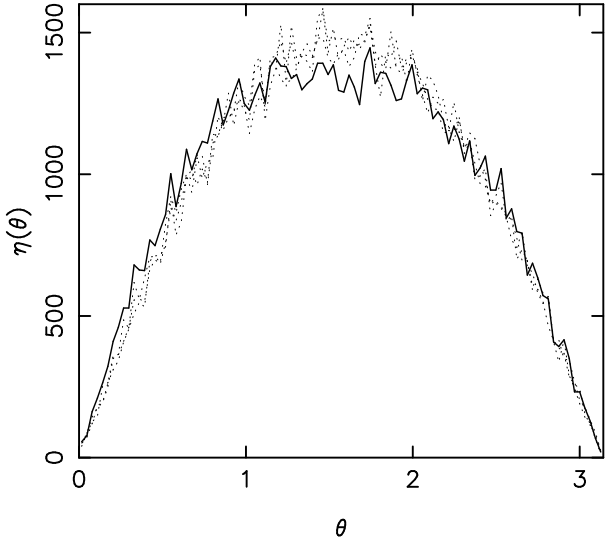} \caption{\label{fig:neg}
Distribution of \(\eta(\theta)\) with $\theta$ for sources with
negative rotation measures. The vertical axis is dimensionless and
the horizontal axis is $\theta$ in radians. The solid line
corresponds to the Broten et al. distribution, and the dotted
lines show three randomly chosen MC distributions. Note the slight
but statistically significant excess at small $\theta$.}
\end{figure}

The distribution  \(\eta^-(\theta)\) calculated from the Broten et
al. catalogue and sample Monte Carlo simulations are shown in
Fig.\ref{fig:neg}. The distributions of \(\eta^+(\theta)\) are
not shown, but they behave in a similar fashion. The \(\chi^2\)
values obtained when comparing both subsamples distributions with
the average MC distributions were larger than those obtained by
all 1000 MC subsamples. Indeed, Fig.1 demonstrates that there
are visible differences between the measured distribution and the
MC simulations. This demonstrates that the real data show a
significant correlation between the sign of rotation measure and
the source's spatial location. This fits in with the idea that
there is an asymmetry about the meridian about the Galactic
centre; a dominance of negative rotation measures at \(0 < l <
180^{\mathrm{o}}\) and positive RM at \(180^{\mathrm{o}} < l <
360^{\mathrm{o}}\) (Vall\'ee \& Kronberg 1975; Han et al. 1997).
Furthermore, the average distribution of both subsamples are
identical in shape; the peaks and troughs correspond to the same
angles. This reinforces the view that both distributions yield
similar information. Overall the results support the view that
rotation measures trace the Galactic magnetic field at the angular
location of the source, which is encouraging for this study.

\section{Correlating Rotation Measures with Sky Maps}

In order to look for correlation between the temperature
measurements and the rotation measures at the locations of the
sources we use a non-parametric measure of correlation. The
temperature (\(T_i\)) at the location of source $i$ and the
rotation measure ${\cal R}_i$ of source $i$ are drawn from unknown
probability distributions. However, if the value of each \(T_i\)
is replaced by the value of its rank among all the other \(T_i\)
then the resulting list of numbers will be drawn uniformly from
integers between 1 and the sample size $N$ (which is 674 for the
whole sample). The same procedure is followed for the ${\cal
R}_i$. We are interested in the magnitude of ${\cal R}$ as this is
determined by the magnetic field strength; the sign is irrelevant
to the intensity of emission at that point. If measurements share
the same value then they are assigned the mean of the ranks that
they would have had if their values slightly differed. In all case
the sum of all ranks should be \(N(N+1)/2\), where $N$ is the
number of sources.

The Spearman rank-order correlation coefficient is constructed as
follows. If \(x_i\) is the rank of \(T_i\) among the other
\(T_i\)'s, and \(y_i\)'s is the rank of ${\cal R}_i$ among the
other ${\cal R}_i$. Then the Spearman rank-order correlation
coefficient, $r_s$ is then given by
\begin{equation}
\label{spearman} r_s=\frac{\sum_i
(x_i-\overline{x})(y_i-\overline{y})}{\sqrt{\sum_i
(x_i-\overline{x})^2}\sqrt{\sum_i (y_i-\overline{y})^2}}
\end{equation}
A perfect positive correlation is represented by $r_s=1$, whereas
$r_s=-1$  is a perfect negative correlation. This is a better
statistic than the usual product-moment correlation coefficient
because there is no reason to suppose a linear relation between
the two variables. In more general terms it is worth stressing
that the non-parametric nature of the Spearman test renders it
insensitive to highly skewed distributions. What it measures
relates to the ordering of the measurements rather than their
actual values so the shapes of marginal distributions of $x$ and
$y$ are irrelevant.

In order to establish the significance of a non-zero \(r_s\) value
obtained from comparing the two sets of measurements, the value
was compared to those obtained through MC simulations. The
simulations were designed such that the temperature field at a
particular point was not linked to the rotation measures. However,
the sources could not be simply placed randomly in any location
the sky  since the real data is clustered which may affect the
significance level of any result. There are two ways this could
have been simulated to take this into account: the rotation
measures of the sources could be shuffled, or the coordinate frame
of the source positions could be rotated. The former option was
chosen as it is computationally faster. We performed 10,000 MCs
and a value of \(r_s\) was obtained for each one. This allows us
straightforwardly to establish the significance level of any
measurement in terms of the fraction of simulated \(r_s\) values
exceeded by the real measurement.

\section{Application to Sky Maps}
\subsection{Foreground Maps} The idea of using rotation measures
to hunt for Galactic contamination in CMB data hinges on the
assumption that foregrounds are related (directly or indirectly)
with the Galactic magnetic field. As we explained above in Section
2, this assumption seems theoretically well-founded but not
empirically proven. If the two sets of measurements were found to
be uncorrelated, it could be that our understanding of the physics
of foreground emission needs adjusting or that the belief that the
the main contribution to the integral in equation (\ref{rm})
relates to the Galactic magnetic field might be incorrect. In
order to verify that there is indeed a correlation between
synchrotron emission and rotation measures we looked at a number
of ``foreground'' maps.

To start with, we looked at 8 ``pure'' synchrotron maps: three of
these are derived from the 408 MHz survey of Haslam et al. (1982)
and 5 are produced by the WMAP team (Bennett et al. 2003); maps
for the latter (and of the data we use later in this section) are
available in HEALPix\footnote{http://www.eso.org/science/healpix/}
format (Gorski, Hivon \& Wandelt 1999) from the NASA archive at
http://lambda.gsfc.nasa.gov. The WMAP maps have a resolution
parameter of \(N_{\mathrm{side}}=256\) corresponding to \(12
\times 256^2= 786,432\) pixels. The 408 MHz map has a resolution
of 51 arcmin and has been reproduced in HEALPix format with
\(N_{side}\)=128. We also looked at this survey smoothed to COBE
resolution (\(N_{side}\)=32). The WMAP team produced their maps
using a maximum entropy method (MEM) with the 408 MHz map
furnishing a prior spatial distribution; the details of the method
are described in Bennett et al. (2003).

\begin{table}\centering
\begin{tabular}{l|c|c}\hline\hline
\emph{Map} & \emph{\(r_s\)}&\emph{Correlation Probability} \\
\hline Synchrotron &&\\ &&\\ 408 MHz (Unmasked) & 0.267 & 1.00\\
408 MHz (COBE resolution) & 0.277 & 1.00\\ 408 MHz (Masked) &
0.237 & 1.0o\\ &&\\ K derived & 0.289 & 1.00\\ Ka derived  & 0.261
& 1.00\\ Q derived & 0.196 & 1.00\\ V derived & 0.168 & 1.00\\ W
derived synchrotron map & 0.094 & 0.99\\ &&\\ Free-Free&&\\ &\\
H$\alpha$ (Unmasked) & 0.444&1.00\\ H$\alpha$ (Masked) &
0.496&1.00\\ WMAP MEM (Unmasked) & 0.427 & 1.00\\ &&\\ Dust&&\\
&\\ FDS model (Unmasked) & 0.405&1.00\\ FDS model (Masked) & 0.429
&1.00\\ WMAP MEM (Unmasked) & 0.247 & 1.00\\
 \hline\hline
\end{tabular}
\caption{ \label{tab:RMvsSYNC} Significance of correlations of
Galactic foreground synchrotron, free--free and dust maps
described in the text with the Broten et al. Faraday rotation
measures.}
\end{table}

The first 8 rows of Table \ref{tab:RMvsSYNC} summarise the results
from these maps. The method identifies weak but highly significant
correlations between the rotation measures and  the 8 maps. The
WMAP-derived maps yield particularly strong signals, with measured
\(r_s\) values exceeding those in all 10,000 MCs except for the
W-band derived case. These strongly significant correlations add
weight to the belief of the WMAP team that they have produced
synchrotron-only maps for the 5 frequencies. While the 408 MHz map
is at such a low frequency that synchrotron emission completely
dominates the sky, the WMAP  maps are produced at frequencies
where the CMB is significant. Furthermore, these maps have been
produced by trying to model and remove other sources of
contamination (dust and free--free). The \(r_s\) values decrease
with frequency across the WMAP range. This is not entirely
unexpected, because synchrotron emission (4) decreases with
frequency. On the other hand simply boosting or suppressing the
amplitude should not influence the non-parametric correlation if
the spatial distribution did not change with frequency. This
effect may therefore be an artifact of variations in $\alpha$.

To explore the possible relationship between free--free radiation
and RM values we used two maps: one by Finkbeiner (2003),
which is a composite all-sky H$\alpha$ map using data from various
surveys, and the other produced by the WMAP team at the K-band
frequency using the MEM approach mentioned above. Perhaps
surprisingly, the results in Table \ref{tab:RMvsSYNC} demonstrate
a stronger correlation between these maps and RM than there is in
the synchrotron maps, suggesting that the correlation may relate
to regions of where the Galactic magnetic field is largely
constant or varies parallel to the line-of-sight but in which the
electron density fluctuates. Free-free radiation will depend on
the square of the electron density so fluctuations in $n_e$ in a
region with constant $B$ may produce larger of correlations of RM
of free--free emission than synchrotron.

As far as dust is concerned we again use two maps: one is a map
constructed by the WMAP team at 94 GHz using models produced by
Finkbeiner, Davis \& Schlegel (1999) and the other is produced by
the WMAP team using MEM at the W-band frequency. Once again the
resulting correlations are significantly stronger than those
displayed by the synchrotron maps.

One possible explanation of the free--free and dust results is
that the cross-correlation with RM we are seeing is basically an
effect of Galactic latitude. If all sources (and $B$) vary with
Galactic latitude then they will correlate with each other. This
is possible, but one piece of evidence against this explanation is
that if we mask out low Galactic latitudes using the Kp2 mask
advocated by Hinshaw et al. (2003), the results for synchrotron do
not decrease dramatically while those for dust and free--free
actually increase slightly.

As a final comment it is worth noting that we repeated the
cross-correlation analysis of these maps, excluding the sources in
the Broten et al. catalogue that yield very large rotation
measures, i.e. ${\cal R}>300$ rad m$^{-2}$. The resulting values
of $r_s$ were all slightly reduced by a maximum of about $0.04$ in
$r_s$. What is interesting about this result is that it shows that
these sources probably do contain some information about the
Galactic magnetic field because their RM measures are not entirely
intrinsic. If these sources only added noise to the
cross-correlation, as we suggested above that they might, then the
result on the cross-correlation of excluding them would be random
rather than systematic.

\subsection{COBE DMR Data}

The DMR instrument comprised six differential microwave
radiometers: two nearly independent channels, labelled A and B, at
frequencies 31.5, 53 and 90 GHz. We use data from the region
\(|b|>\mathrm{20^{o}}\) with custom cutouts near the Galactic
centre \cite{bgbh97}. There are 460 sources from the Broten et al.
catalogue in this region to compare with the DMR data. The dipole
anisotropy of amplitude \(\sim\) 3 mK is largely removed in
pre-map-making process \cite{bddd03}. We chose to look for
correlations in 15 maps from the DMR data: the raw data from each
of the six radiometers; the (A+B)/2 sum maps and (A-B)/2
difference maps for each frequency; and the (A+B)/2 sum maps after
smoothing with a \(\mathrm{7^{o}}\) beam up to $l$=20 using the
'smoothing' routine in the HEALPix package. The sum maps should
represent the true CMB signal whereas the difference maps should
measure the level of instrument noise. The COBE-DMR four year sky
maps used for this analysis have a resolution parameter of
\(N_{\mathrm{side}}=32\) corresponding to \(12 \times 32^2=
12,288\) pixels in the HEALPix representation.

\begin{table}\centering
\begin{tabular}{l|c|c}\hline\hline
\emph{Map} & \emph{\(r_s\)}&\emph{Correlation Probability} \\
\hline 31.5 GHz  channel A & 0.039 & 0.80\\ 31.5 GHz  channel B &
0.013 & 0.61\\ 53.0 GHz  channel A & -0.074 & 0.06\\ 53.0 GHz
channel B & -0.024 & 0.29\\ 90.0 GHz  channel A & 0.062 & 0.91\\
90.0 GHz  channel B & -0.031 & 0.26\\ &\\ 31.5 GHz (A-B)/2 & 0.006
& 0.56\\ 53.0 GHz (A-B)/2 & -0.026 & 0.29\\ 90.0 GHz (A-B)/2 &
0.059 & 0.90\\ &\\ 31.5 GHz (A+B)/2 & 0.017 & 0.64\\ 53.0 GHz
(A+B)/2 & -0.045 & 0.17\\ 90.0 GHz (A+B)/2 & 0.045 &0.83\\ &\\
31.5 GHz (A+B)/2 smoothed & 0.056 & 0.89\\ 53.0 GHz (A+B)/2
smoothed & 0.002 & 0.51\\ 90.0 GHz (A+B)/2 smoothed & -0.048 &
0.15\\\hline\hline
\end{tabular}
\caption{ \label{tab:RMvsCOBE} Significance levels of correlations
between the DMR maps described in the text and the  Broten et al.
data.}
\end{table}

The results of looking for correlations between the DMR maps and
the rotation measures are shown in table \ref{tab:RMvsCOBE}. The
sum maps (corresponding to CMB + contaminants) should show the
strongest signs of correlations if there are any to be seen.
However, none of the maps show any evidence of correlation with
the Broten et al. catalogue. In order to see how stable the
probability values are, the simulations were repeated for the 90
GHz (A+B)/2 smoothed map a further three times giving
probabilities of 0.16, 0.15 and 0.15. This indicates that the
probabilities and thus any conclusions drawn from them are valid.
Although the main factor may well be the relatively low
signal-to-noise in the COBE-DMR data, we can say that our method
shows no evidence for any residual foreground component in these
maps.

\subsection{WMAP 1 yr Sky Maps}

The WMAP instrument comprises 10 differencing assemblies
(consisting of two radiometers each) measuring over 5 frequencies
(\(\sim \)23, 33, 41, 61 and 94 GHz). The two lowest frequency
bands (K and Ka) are primarily Galactic foreground monitors, while
the highest three (Q, V and W) are primarily cosmological bands
\cite{hsvh03}. For CMB analyses, it is necessary to mask out
regions of strong foreground emission. Bennett et al. (2003; B03)
provide masks for excluding regions where the contamination level
is large. The masks are based on the K-band measurements, where
contamination is most severe. The masks, of differing levels of
severity, are available from the NASA archive. The severity of the
mask is a compromise between eliminating foregrounds and
maximising sky area in analyses. We chose to use the Kp2 mask used
by Hinshaw et al. (2003) to calculate cross-spectra from the three
high frequency data leading to the angular power spectrum. The
mask removes 15 \% of pixels (including bright sources) leaving
338 faraday sources to compare with the maps. There should be no
correlations between the rotation measures and the high frequency
maps, once the mask has been applied. If there are correlations,
this would question whether the amplitude of the angular power
spectrum is cosmological in origin.

The WMAP team have also released an internal linear combination
(ILC)  map that combined the five band maps in such a way to
maintain unity response to the CMB whilst minimising foreground
contamination. The construction of this map is described in detail
in Bennett et al. (2003). To further improve the result, the
inner Galactic plane is divided into 11 separate regions and
weights determined separately. This takes account of the spatial
variations in the foreground properties. Thus, the final combined
map does not rely on models of foreground emission and therefore
any systematic or calibration errors of other experiments do not
enter the problem. The final map covers the full-sky and should
represent only the CMB signal.

Following the release of the WMAP 1 yr data Tegmark, de
Oliveira-Costa \& Hamilton (2003; TOH) have produced a cleaned CMB
map. They argued that their version contained less contamination
outside the Galactic plane compared with the internal linear
combination map produced by the WMAP team. The five band maps are
combined with weights depending both on angular scale and on the
distance from the Galactic plane. The angular scale dependence
allows for the way foregrounds are important on large scales
whereas detector noise becomes important on smaller scales. TOH
also produced a Wiener filtered map of the CMB that minimises
rms errors in the CMB. Features with a high signal-to-noise are
left unaffected, whereas statistically less significant parts are
suppressed. While their cleaned map contains residual Galactic
fluctuations on very small angular scales probed only by the W
band, these fluctuations vanish in the filtered map.

In all, 13 maps derived from the WMAP data were used to seek
correlation with the Broten et al. catalogue: the five band maps,
the five band maps with the Kp2 region removed, the internal
linear combination map, and the cleaned and Wiener maps of TOH.

\begin{table}\centering
\begin{tabular}{l|c|c}\hline\hline
\emph{Map} & \emph{\(r_s\)}&\emph{Correlation Probability} \\
\hline K & 0.292 & 1.00\\ Ka & 0.189 & 1.00\\ Q & 0.149 & 1.00\\ V
& 0.126 & 1.00\\ W & 0.063 & 0.95\\ &\\ K (with mask) & 0.296 &
1.00\\ Ka (with mask) & 0.099 & 0.97\\ Q (with mask) & 0.041 &
0.77\\ V (with mask) & 0.049 & 0.81\\ W (with mask) & 0.043 &
0.78\\ &\\ Internal linear & 0.021 & 0.71\\ combination map&&\\
&\\ TOH cleaned map  & 0.059 & 0.93\\ TOH Wiener map & 0.068
&0.96\\\hline\hline
\end{tabular}
\caption{ \label{tab:RMvsWMAP} Significance levels of
cross-correlations  derived from various maps originating from
WMAP with the Broten et al. catalogue.}
\end{table}

The results of looking for correlations between the WMAP derived
data and the rotation measures are shown in Table
\ref{tab:RMvsWMAP}. The uncut frequency maps are all significantly
correlated with the rotation measures, except the highest
frequency map (W band). This confirms the expected contamination
of the data across these frequencies. The strength of correlation
decreases with increasing frequency, which may
be understood by looking at the contribution of all the foregrounds to the
total sky signal across these bands. Combining the contributions of
synchrotron, free--free and dust, with sychrotron given a weighting of 0.5
due to its weaker correlation, would reproduce this trend. Once the Kp2 mask has
been applied, the correlations vanish for all but the K and Ka
band map. The reason that a correlation is still found with the K
band data is probably that the cut only excludes the very
strongest signals so both bands are still heavily contaminated.
This is probably not a problem for CMB analysis because these will
be projected out in the likelihood analysis.  Moreover, the
results from the masked maps indicate that calculating the angular
power spectrum from the 3 high frequency bands with the Kp2 cut is
valid. The contamination from Galactic foregrounds correlated with
Faraday rotation on these studies is therefore probably small.
Finally, no correlations are found between the rotation measures
and two of the CMB-only maps, suggesting that the levels of
contamination are indeed low, as the authors claim. Nevertheless,
the result from the Wiener filtered map of TOH suggests a
correlation at 95\% confidence level.

\begin{table}\centering
\begin{tabular}{l|c|c}\hline\hline
\emph{Map} & \emph{\(r_s\)}&\emph{Correlation Probability}
\\\hline q1 cleaned  &0.005  &0.53\\ q2 cleaned  &-0.075
&0.08\\ &\\ v1 cleaned  &-0.008 &0.43\\ v2 cleaned  &0.023 &0.66\\
&\\ w1 cleaned  &-0.048  &0.19\\ w2 cleaned  &0.009 &0.56\\ w3
cleaned  &-0.047  &0.19\\ w4 cleaned  &0.018 &0.63\\\hline\hline
\end{tabular}
\caption{Cross-correlations in the ``clean maps'' from each high
frequency assembly}
\end{table}

Finally, we mention that on the NASA archive
 there are maps from each high
frequency assembly that are supposedly clean of foreground
contaminants outside the Kp2 cut region. The results in Table 4
show there are indeed no significant residual correlations in
these data when the Kp2 mask is applied.

\section{Discussion and Conclusions}

The aim of the paper was to look for traces of Galactic
contamination in the COBE-DMR 4 yr data and WMAP 1 yr data using
correlations  between the maps and the rotation measures of Broten
et al. (1988).

We first studied the relationship between the spatial position and
the RM values of the sources in the catalogue by looking at the
angular correlations of subsets drawn from it. The results clearly
indicate a correlation,  and confirm the basic view expressed by
other authors on properties of the Galactic magnetic field. We
then investigated the relationship between the rotation measure of
a source and the synchrotron emission at the location of the
source. Synchrotron maps derived from the 408 MHz survey of Haslam
et al. (1982) and produced by the WMAP team were found to be
correlated with RM values, showing that these do indeed provide
some sort of probe of the Galactic magnetic field.

We found {\em stronger} positive correlations of RM values with
both dust and free--free maps than for synchrotron. This is
consistent with indirect association of the different sources, but
is probably not simply a Galactic latitude effect because the
correlation persists even when a Galactic cut is applied. Exactly
how this correlation arises is an issue for further study, but it
may provide an insight into the possible role of spinning dust
(Draine \& Lazarian 1998) in Galactic foregrounds \cite{spin} as
this may align with the local magnetic field. Clearly much more
detailed modelling is needed to understand these correlations
theoretically, but the empirical approach we have adopted still
provides a useful consistency check on foreground analysis even if
the origin of the correlation is not well understood.

The correlations between the temperature strength of the DMR and
WMAP data at the location of sources and the RM values were then
studied using the Spearman rank-order correlation coefficient. All
15 maps compiled from the temperature field measured by the DMR
instrument were found to be uncorrelated with the rotation
measurements. Correlations were found with the uncut WMAP
frequency maps and the cut K and Ka band maps. However, the maps
used by the WMAP team to extract cosmological information were
found to be uncorrelated. Furthermore, two foreground-subtracted
CMB-only maps were found to be uncorrelated with the rotation
measure catalogue. The results of this analysis provide no
evidence of residual foregrounds in the COBE-DMR maps or the WMAP
ILC map, but do yield a positive correlation for the Tegmark
Wiener-filtered map. Owing to the small size of our RM sample we
used these results are only suggestive, but they do demonstrate
the virtue of looking for independent probes of Galactic
foreground contamination. Much larger compilations of RM values
would be needed to make more definite statements about
contamination in temperature maps. It is also very
likely that much could be learnt about polarized foregrounds
using a similar approach.

\section*{Acknowledgements}
We thank Anvar Shukurov for sending us the rotation measure
catalogue used for this work and Tony Banday for supplying us with
the COBE-DMR data and synchrotron maps in a convenient format. We
gratefully acknowledge use of the HEALPix package and the Legacy
Archive for Microwave Background Data Analysis (LAMBDA). Support
for LAMBDA is provided by the NASA Office of Space Science.
Finally, we are immensely grateful to an anonymous referee who
provided us with many helpful suggestions which enabled us to
improve the paper substantially.

\end{document}